\documentclass[apjl]{emulateapj}
\usepackage{xcolor}

\usepackage{natbib}

\shorttitle{GW150914: First TOROS search for GW-EM counterpart}
\shortauthors{Diaz, M. et al.}
\begin{document}

\title{GW150914: First search for the electromagnetic counterpart of a gravitational-wave event by the TOROS collaboration}

\author{Mario C.~D\'{\i}az\altaffilmark{*,1}, Mart\'{\i}n Beroiz\altaffilmark{1,2}, Tania Pe\~{n}uela\altaffilmark{1,3}, Lucas M.~Macri\altaffilmark{4}, Ryan J.~Oelkers\altaffilmark{4,10}, 
Wenlong Yuan\altaffilmark{4}, Diego Garc\'{\i}a Lambas\altaffilmark{5}, Juan Cabral\altaffilmark{5}, Carlos Colazo\altaffilmark{5, 6}, Mariano Dom\'{\i}nguez\altaffilmark{5}, Bruno S\'anchez\altaffilmark{5}, Sebasti\'an Gurovich\altaffilmark{5}, Marcelo Lares\altaffilmark{5}, Mat\'{\i}as Schneiter\altaffilmark{5}, Dar\'{\i}o Gra\~{n}a\altaffilmark{5}, V\'{\i}ctor Renzi\altaffilmark{5}, Horacio Rodriguez\altaffilmark{5},  Manuel Starck\altaffilmark{5}, Rub\'en Vrech\altaffilmark{5}, Rodolfo Artola\altaffilmark{5},  Antonio Chiavassa Ferreyra\altaffilmark{5},  Carla Girardini\altaffilmark{5}, Cecilia Qui\~{n}ones\altaffilmark{5},  Luis Tapia\altaffilmark{5}, 
Marina Tornatore\altaffilmark{5}, Jennifer L.~Marshall\altaffilmark{4}, Darren L.~DePoy\altaffilmark{4}, Marica Branchesi\altaffilmark{7}, Enzo Brocato\altaffilmark{8}, Nelson Padilla\altaffilmark{9}, Nicolas A.~Pereyra\altaffilmark{1}, Soma Mukherjee\altaffilmark{1}, Matthew Benacquista\altaffilmark{1} \& Joey Key\altaffilmark{1}}
\altaffiltext{*}{Corresponding author, mario.diaz@utrgv.edu}
\altaffiltext{1}{Center for Gravitational Wave Astronomy, University of Texas Rio Grande Valley, Brownsville, TX, USA}
\altaffiltext{2}{University of Texas at San Antonio, San Antonio, TX, USA}
\altaffiltext{3}{Ludwig Maximilian Universit\"at Munich, Faculty of Physics, Munich, Germany}
\altaffiltext{4}{Mitchell Institute for Fundamental Physics \& Astronomy, Department of Physics \& Astronomy, Texas A\&M University, College Station, TX, USA}
\altaffiltext{5}{Universidad Nacional de C\'ordoba, IATE, C\'ordoba, Argentina}
\altaffiltext{6}{Ministerio de Educaci\'on de la Provincia de C\'ordoba, Argentina}
\altaffiltext{7}{Universit\`a degli studi di Urbino, Urbino, Italy}
\altaffiltext{8}{INAF - Osservatorio Astronomico di Roma, Monte Porzio Catone, Italy}
\altaffiltext{9}{Instituto de Astrof\'isica, Pontificia Universidad Cat\'olica de Chile, Santiago, Chile}
\altaffiltext{10}{Current address: Physics \& Astronomy Department, Vanderbilt University, Nashville, TN, USA}
\begin{abstract}
We present the results of the optical follow-up conducted by the TOROS collaboration of the first gravitational-wave event GW150914. We conducted unfiltered CCD observations ($0.35-1\mu$m) with the 1.5-m telescope at Bosque Alegre starting $\sim2.5$ days after the alarm. Given our limited field of view ($\sim100\sq\arcmin$), we targeted 14 nearby galaxies that were observable from the site and were located within the area of higher localization probability.

We analyzed the observations using two independent implementations of difference-imaging algorithms, followed by a Random-Forest-based algorithm to discriminate between real and bogus transients. We did not find any {\it bona fide} transient event in the surveyed area down to a $5\sigma$ limiting magnitude of $r=21.7$~mag (AB). Our result is consistent with the LIGO detection of a binary black hole merger, for which no electromagnetic counterparts are expected, and with the expected rates of other astrophysical transients.
\end{abstract}

\keywords{Gravitational Waves, General relativity, GW150914, techniques: image processing}

\section{Introduction}
\setcounter{footnote}{10}
The network of advanced ground-based gravitational wave (GW) interferometers constituted by the LIGO observatories \citep{LSC2015}, which started operations in September 2015 and by the VIRGO observatory \citep{Acernese2015}, which will join before the end of 2016, were designed to be capable of detecting GWs emitted by the mergers of neutron stars and/or black holes in binary systems out to distances of hundreds of Mpc \citep[see][and references therein]{Abbott2016b}. In anticipation of the operation of this network, on 2013 June 6 the LIGO-VIRGO collaboration (LVC) issued a worldwide call\footnote{http://www.ligo.org/scientists/GWEMalerts.php} to astronomers to participate in multi-messenger observations of astrophysical events recorded by the GW detectors, using a wide range of telescopes and instruments of ``mainstream'' astronomy.

\ \par

Initially, triggers will be shared promptly only with astronomy partners who have signed a Memorandum of Understanding (MoU) with LVC involving an agreement on deliverables, publication policies, confidentiality, and reporting. It is expected that if the mergers of compact objects contain at least one neutron star, electromagnetic (EM) radiation will be emitted during the event. This EM counterpart, originating in the ejecta and its interaction with the surrounding environment could range from very short duration gamma-ray bursts to longer-duration emission at optical, near infrared (kilonova and short GRB afterglows) and radio wavelengths \citep[e.g.,][]{Li1998,Nakar2011,Metzger2012,Barnes2013,Berger2014,Cowperthwaite2015}. Simultaneous detection of the event by GW and EM observatories could provide a more integrated  astrophysical interpretation of the event and would be instrumental in producing better estimates for the distance and energy scales of the event.

Motivated to participate in these observations, we formed a collaboration named ``Transient Optical Robotic Observatory of the South'' \citep[TOROS;][]{Benacquista2014} which seeks to deploy a wide-field optical telescope on Cord\'on Mac\'on in the Atacama Plateau of northwestern Argentina \citep{Renzi2009,Tremblin2012}. The collaboration planned to utilize other resources independently of the construction of this facility. On 2014 April 5, the TOROS collaboration signed a Memorandum of Understanding with LVC  and participated during the first scientific run of the GW interferometers from September 2015 through January 2016. Two facilities were available to TOROS during this campaign: a Schmidt-Cassegrain 0.4-m telescope at Cord\'on Mac\'on and a 1.5-m telescope at Estaci\'on Astrof\'{\i}sica Bosque Alegre (EABA) in C\'ordoba, Argentina.  

On 2015 September 14 at 09:50:45 UT, the two USA-based detectors of the Advanced LIGO interferometer network detected a high-significance candidate GW event designated GW150914 \citep{Abbott2016a}. This unexpected detection --- observed four days before the first scientific run of the detectors was scheduled to start --- constituted the first detection of the merger of a binary black hole (BBH) system and the first direct detection of gravitational waves. Due to the unexpected timing of the event, LVC provided spatial location information two days later, in the form of probability sky maps via a private GCN circular \citep[GCN\#18330]{Singer2015}. TOROS was one of 25 teams that participated in the search for an electromagnetic counterpart in the southern hemisphere. We report here on the optical follow up of this event by the TOROS collaboration during 2015 September 16-17 using the 1.5-m EABA telescope (the smaller telescope at Mac\'on was not operational at the time).

This paper is organized as follows: \S\ref{observations} discusses target selection and observations; \S\ref{data-analysis} describes the data reduction, image differencing algorithms and the bogus/real classification; \S\ref{results} presents our results and \S\ref{conclusions} summarizes our findings. Throughout this paper, we express magnitudes in the AB system and adopt a $\Lambda$CDM cosmology based on results from the Planck mission \citep{Planck2015}.

\section{Observations}\label{observations}

\begin{deluxetable*}{clrrrrrr}[h!]
\tablecaption{Targeted host galaxies\label{table1}}
\tablehead{ \colhead{Date$^{1}$} & \colhead{ID$^{2}$} & \colhead{RA} & \colhead{Dec.} & \colhead{$t_{\rm exp}$} & \colhead{Tile \#} & \colhead{D}\\
\colhead{} & \colhead{} & \multicolumn{2}{c}{[deg]} & \colhead{[s]} & \colhead{} & \colhead{[Mpc]}}
\startdata
2015-09-16& IC1933    &  51.416101& -52.78547&  600& 1,2,3,4   & 17.45&  \\
2015-09-16& NGC1529   &  61.833301& -62.89993&  600& 5,6,7,8   & 54.76&  \\
2015-09-16& IC2038    &  62.225246& -55.99074&  600& 9,10,11,12&  7.00& \\
2015-09-16& IC2039    &  62.259901& -56.01172&  600& 9,10,11,12&  7.63&  \\
2015-09-17& ESO058-018& 102.593850& -71.03123& 1020& 13        & 52.23&  \\
2015-09-17& ESO084-015&  65.550449& -63.61097& 1140& 14        & 14.99&  \\
2015-09-17& ESO119-005&  72.072451& -60.29376& 1080& 15        &  9.73&  \\
2015-09-17& NGC1559   &  64.398901& -62.78358&  900& 16        & 12.59&  \\
2015-09-17& PGC016318 &  73.728898& -61.56747& 1020& 17        &  9.54&  \\
2015-09-17& PGC269445 & 100.209150& -71.33026& 1140& 18        & 54.83& \\
2015-09-17& PGC280995 &  96.382499& -69.15257& 1140& 19        & 55.08&  \\
2015-09-17& PGC128075 &  64.859998& -60.53844&  720& 20        & 63.71&  \\
2015-09-17& PGC381152 &  63.584547& -58.20726& 1200& 21        & 13.26& \\
2015-09-17& PGC075108 &  63.670349& -58.13199& 1200& 21        & 13.29& 
\enddata
\tablecomments{(1) local date of observation; (2) from \citet{White2011}}
\end{deluxetable*}

For the first run of the LIGO detectors (O1) several low-latency analyses were prepared to receive and process signals from GWs. On 2015 September 16, the LIGO Virgo Collaboration (LVC) provided two all-sky localization probability maps for the event, based on them. Both the coherent Wave Burst \citep[cWB;][]{Klimenko2016} and the Omicron+LALInference Burst \citep[oLIB;][]{Lynch2015} search for un-modeled signals. The first one, a rapid localization analysis just searches for coherent power across both detectors while the second one, more refined, assumes a Sine-Gaussian content. The maps provided initial spatial localization of 50\% and 90\% confidence regions encompassing about 200 and 750 $\sq^{\circ}$, respectively  \citep[][GCN\#18330]{Singer2015}.

We started our imaging campaign immediately after receiving the GCN circular and acquired the first epoch of observations on 2015 September 16 \& 17. We obtained a second epoch of imaging (to serve as templates for the differencing pipelines) on 2015 December 5 \& 6. We used an Apogee Alta U9 camera with a field of view (FoV) of $12\farcm 7 \times 8\farcm 5$ and an effective plate scale of $0\farcs 75 \  {\rm pix}^{-1}$ after $3\times 3$ binning. Since we wished to maximize our sensitivity, we conducted unfiltered (``white light'') observations spanning $0.35 < \lambda/\mu m < 1$. We obtained individual exposures of 60~s with a median seeing (FWHM) of $(2.8\pm0.6)\arcsec$. We typically obtained 10 images per field, reaching $5\sigma$ limiting magnitudes of $r=21.7\pm0.3$~mag (see \S\ref{data-analysis} for details).

\begin{figure*}[t]
\centering

\includegraphics[scale=0.80]{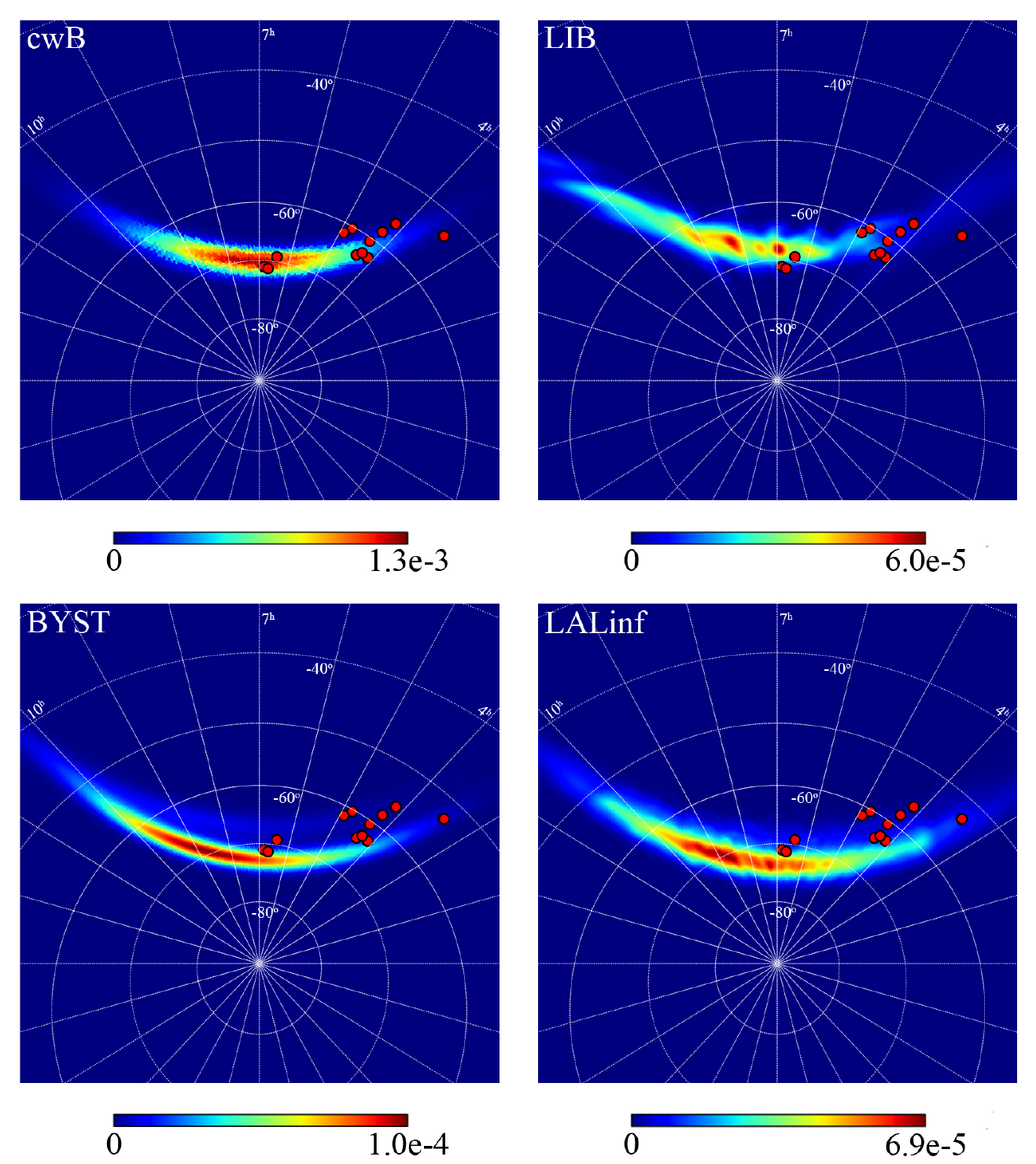}

\caption{Localization probability maps for GW150914 generated by various LIGO pipelines, indicating the location of the TOROS targets (red dots). \label{figmap}}
\end{figure*}

The LIGO localization regions span several hundred square degrees (see Fig.~\ref{figmap}) and vary depending on the algorithm. For instance, the $90\%$ credible localization area for cWB covers to $310\sq^{\circ}$ while others span up to $750\sq^{\circ}$ \citep[see table 1 in][]{Abbott2016d}. Regardless, all sky maps are consistent with a broad long arc in the Southern hemisphere and a smaller extension in the Northern hemisphere.  The algorithm utilized for the CWB estimations produces reasonably accurate maps for BBH signals, but underestimates the extent of high-confidence regions \citep{Essick2015}. As seen in Fig.~\ref{figmap}, the adoption of maps from alternative algorithms (not available at the time our observations started) significantly reduces the fraction of the high-confidence region probed by our small FoV.

Previous work in the field \citep{Nuttall2010,Abadie2012,Hanna2014} have shown that using a galaxy catalog can greatly increase the probability of finding an EM counterpart in the case of BNS or NSBH events. As the LIGO analysis was still ongoing at the time our observations had begun and the nature of the binary was unknown, we optimized the use of our small FoV by targeting nearby galaxies with the highest probability of hosting the event. The probabilities were based on the values of the pixel in the initial cWB map that contained the coordinates of a given galaxy. We used the Gravitational Wave Galaxy Catalog \citep[GWGC;][]{White2011}, which is a compilation of catalogs homogenized into a list of $\sim53000$ galaxies within 100 Mpc (with incompleteness starting at $D\sim40$~Mpc). The GWGC provides reliable distances, blue magnitudes and other properties. 

Table~\ref{table1} lists the galaxies targeted in our search. They were selected using an in-house ``scheduler'' (a Python module of the TOROS pipeline). The scheduler set a list of criteria: (1) observability from our location ($30^\circ > \delta > -70^\circ$), (2) apparent magnitude $B \leq 21$~mag, and (3) distance $D<60$~Mpc.  We plan to add for future observations absolute magnitude $M_B \leq -21$~mag as an additional criterion.  This cut in absolute magnitude is motivated by the expectation that in the nearby Universe the distribution of BNS and BHs should follow recent star formation due to the short merger timescales \citep[see e.g. ][]{Phinney1991,Belczynski2002}. Once we cross-matched the LIGO sky maps with the filtered galaxies, we ranked the results by assigning individual probabilities $P_{g,i}$ (with $i$ being the sky map pixel that contained the $g$-th galaxy). This enabled us to prioritize targets according to their location within the sky maps and their observability. Lastly, we ensured that all targets were mapped out to $\sim5$kpc, which corresponds to the median offset distance of short GRBs from hosts galaxies measured from the optical afterglow observations \citep{Church2011,Fong2013,Berger2014}. This required tiling to cover the appropriate area for some targets. A total of 21 fields covering 14 galaxies were observed. These correspond to $\sim4.4\%$ of the potential host galaxies listed in the GWGC that met selection criteria (2) \& (3). We note that at $D\sim60$~Mpc, the GWGC is estimated to be complete at the $\sim80$\% level \citep{White2011}. 

\ \par 

\section{Data analysis}\label{data-analysis}
The initial data reduction followed the standard steps of bias and dark subtraction, flat-fielding using twilight sky frames, and illumination correction, based on common routines available in PyRAF and independent Python modules that constitute the TOROS data processing pipeline.  Astrometric solutions were derived using the {\it Astrometry} package \citep{Lang2009}, a very robust algorithm based on {\it geometrical~hashing} of asterisms and Bayesian decision trees that uses all-sky catalogs such as USNO-B \citep{Monet2003}, 2MASS \citep{Skrutskie2006} and {\it GALEX} \citep{Martin2003}.

\ \par

Flux calibration was obtained by performing aperture photometry with DAOPHOT \citep{Stetson1987} and matching the resulting star lists against the APASS catalog \citep{Henden2016}; we found 208 stars in common with $0.4 < B-i < 3.2$. The photometric solution was based on the $r$ band since it exhibited the smallest color term for our unfiltered observations, yielding a zeropoint uncertainty of 0.054~mag. We used the reported photometric errors from DAOPHOT and our photometric calibration to estimate a median $5\sigma$ limiting magnitude of $r=21.7\pm0.3$~mag for our fields.

\ \par

\subsection{Difference Imaging  Analysis} \label{2kk}
We carried out two independent implementations of difference-imaging analysis (DIA) to identify transients. Most DIA routines use a kernel defined as a combination of two or more Gaussians to match and scale the point-spread-function (PSF) between two epochs \citep{Alard1998}, which leads to difficulties in fitting irregular shaped PSFs. Our implementations go beyond this simple approach.

Our first method (hereafter ``Method I''), described in \citet{Oelkers2015}, uses a Dirac $\delta$-function kernel fit across the entire frame. We selected the epoch with the smaller PSF to act as the reference frame. Additionally $13\times13$~pix stamps were taken around isolated stars with photometric precision better than 0.05 mag to solve for the kernel coefficients using the least-squares method. We modeled the spatial variation in the PSF with a $9\times9$ pix first-order kernel \citep{Alard2000,Miller2008} if there were at least 20 stars to solve for the coefficients; otherwise, we adopted a constant PSF.

The second algorithm (hereafter, ``Method II'') relies on an independent pixel-by-pixel fit of the convolving kernel \citep{Bramich2008} with a simultaneous polynomial local background fit on the grid of an image. One of the advantages of this method is that the basis functions are removed, so the user does not need to choose a kernel (which in some cases could become inappropriate). Moreover, the basis functions are constructed around the origin of the kernel coordinate system, which requires a very good  alignment of the images for an optimal result. However, the strongest caveat of this method is the type of  grid used to cover the image, as kernels may change abruptly from site to site, due to the fact that there is no kernel interpolation applied between image subsections. Although an overabundance of spurious subtraction artifacts (``bogus'' detections) is obtained with this method, both methods appear to be very effective and have shown similar results.

\subsection{Real/Bogus classification and\\detection of potential transients} \label{method}

We trained a supervised machine-learning algorithm to discriminate between bona fide astrophysical transients and ``bogus'' detections arising from DIA artifacts.   In order to do so, we injected 100 artificial stars on each of the 21 science images,  repeating the procedure 10 times to improve our statistics, and subjected the resulting 210 frames to the same DIA methods described above.  The injections followed the same magnitude distribution as the point sources detected by DAOPHOT in the reference images.

We ran SExtractor \citep{Bertin1996} on the differenced image products from both methods to detect objects with a significance of $>2\sigma$ and with $>5$ connected pixels and extract their defining characteristics (such as magnitude, magnitude error, ellipticity and sharpness). We identified the objects that corresponded to known injected sources and labeled them accordingly. We used a random-forest algorithm with $5\times$ cross validation to identify any other sources exhibiting properties similar to the injections, as potential astrophysical transients. We rejected all other remaining sources, here-after identified as  ``bogus''.

\section{Results}\label{results}
SExtractor detected $\sim10400$ and $\sim34000$ objects on the 210 frames processed with Methods I \& II, respectively; of these, 5441 and 5824 were recovered artificial stars. We used these recovered injections to define the set of features needed to identify ``real'' detections and remove ``bogus'' candidates. 

\begin{figure}[h!]
\centering
\includegraphics[scale=0.4]{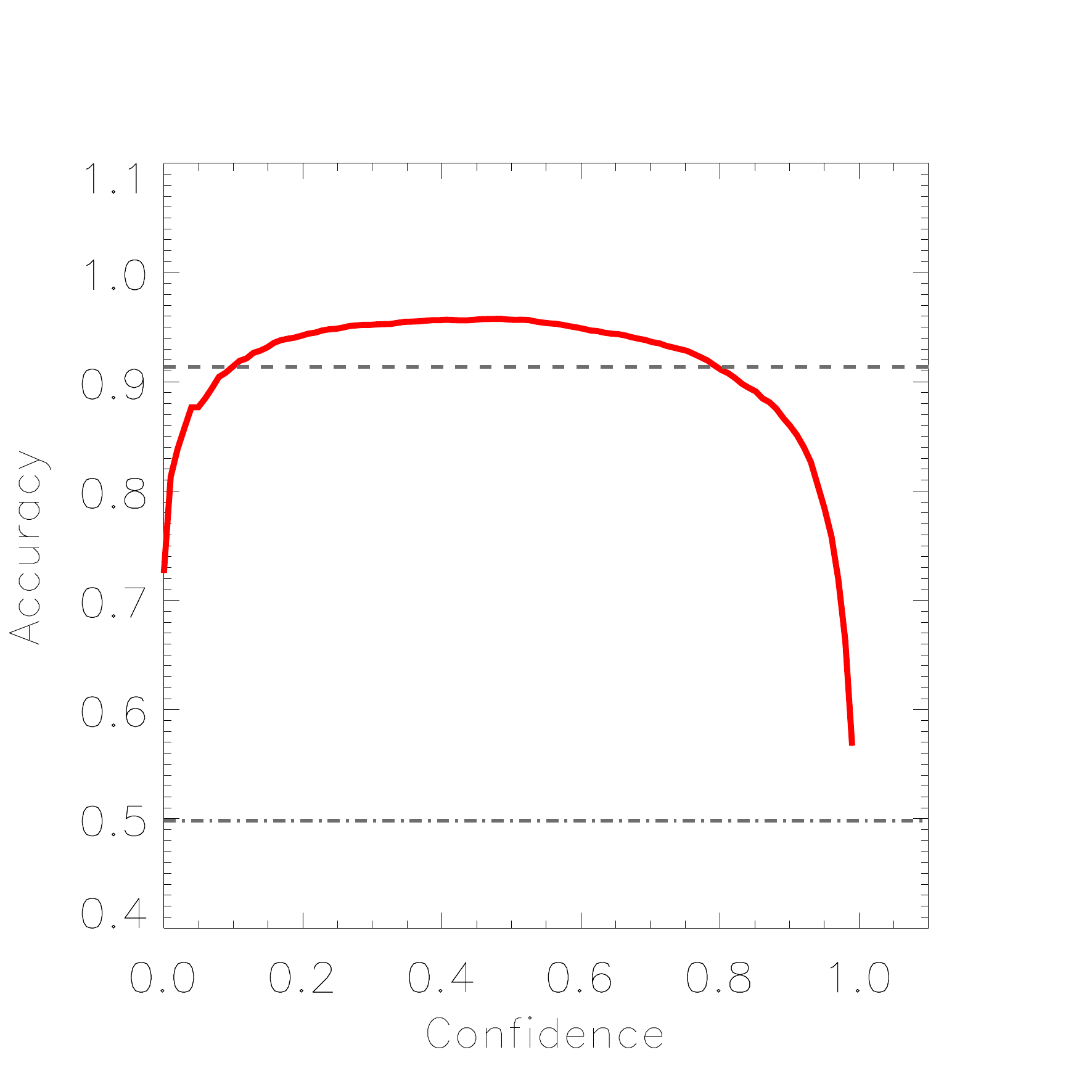} \includegraphics[scale=0.4]{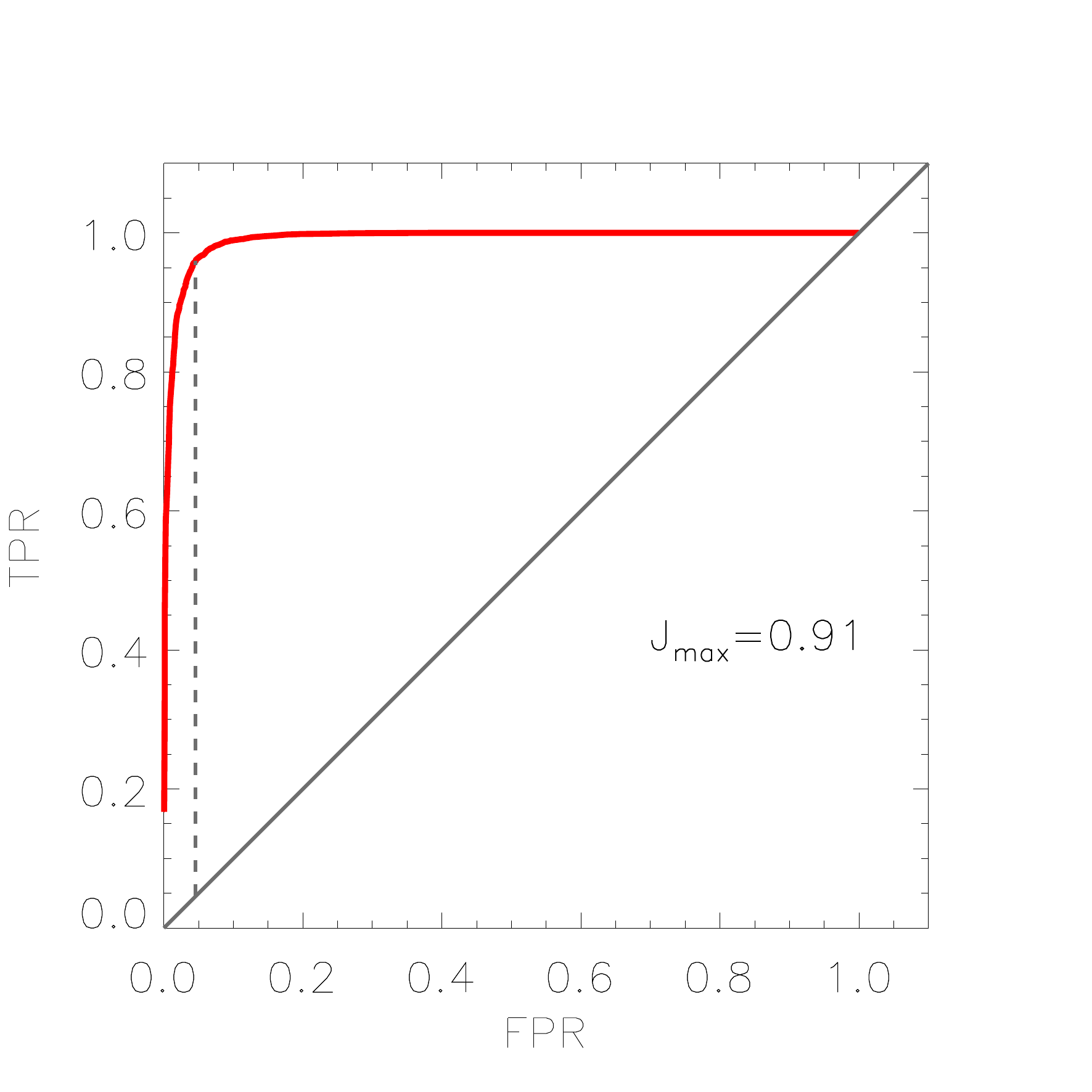}
\caption{Top: {Plot of model accuracy vs.~confidence}. Bottom: {Receiver Operating Curve. See text for details}.\label{figures}}
\end{figure}

Fig.~\ref{figures} shows the results of the Random Forest classification, including model accuracy vs.~confidence and the Receiver Operating Curve (ROC) for the set of transients obtained through Method I (defined in \S3.1). The upper panel also shows the average accuracy (upper dashed line) and the probability of obtaining the same result by chance (dashed-dotted line). The latter is computed using the $P_{e}$ statistic, defined as the sum of the probabilities of the model for either predicting a ``real'' transient ($P_r$) or a ``bogus'' event ($P_b$): $P_e=P_r+P_b$. The respective probabilities are calculated as follows:
\begin{equation}
P_e =\left[ \left( \frac{A_r}{N} \right)  \left( \frac{P_r}{N} \right) + \left( \frac{A_b}{N}\right) \left(\frac{P_b}{N} \right) \right]
\end{equation}
\noindent where $A_r$ and $A_b$ is the number of injections and unknowns and $N$ is the total number of objects in the sample.

We used the ROC to calculate Youden's statistic \citep{Youden1950} or informedness $J_{\rm max}=0.91\pm0.0039$, where the quoted error represents the $95\%$ confidence level \citep{Powers2011}. $J_{\rm max}$ gives the maximum performance of the model and is defined as the maximum distance between the ROC and the 1:1 line that represents the probability of obtaining the result by chance. The observed value of $J_{max}$ corresponds to a cutoff of $0.49$ in terms of the confidence value. It is in good agreement with the maximum accuracy of 0.96 reached by the classifier, as seen in Fig.~\ref{figures}. We therefore selected a cutoff value of 0.5 for our final analysis and the resulting confusion matrix is presented in Table~\ref{conf.matrix}.

\begin{deluxetable}{crr}
\tablecaption{Confusion Matrix at $p=0.5$ \label{conf.matrix}}
\tablehead{            & \multicolumn{2}{c}{Prediction}\\
\colhead{Actual} & \colhead{1} & \colhead{0}}
\startdata
   1& 0.961 & 0.038 \\  
   0& 0.046 & 0.953   
\enddata
\end{deluxetable}

Following these procedures, we identified 229 and 200 objects in all the images processed by Methods I \& II, respectively, as having probabilities greater than 50\% of being real. As a final discrimination against spurious detections, we required an object to be detected in at least 5 of 10 realizations of a given field in order to be considered a \textit{bona fide} astrophysical transient. None of the objects in either set passed this requirement. Further visual inspection revealed most of them to be subtraction residuals or cosmetic defects in the detector. We therefore conclude that no transients were present in the 21 fields we targeted, to a $5\sigma$ limiting magnitude of $r=21.7$ mag. Similar results were obtained for Method II, albeit with a higher fraction of ``bogus'' sources, a result leading to a less balanced sample to merit further analysis. 

The fact that we did not find any genuine transient in our search is not surprising given the small area surveyed ($\sim0.62 \sq^{\circ}$) and the low cadence of the observations (with only two epochs per target, separated by $73\pm3$~d). Based on our temporal sampling and photometric precision, we scaled the results of \citet{Oelkers2015, Oelkers2016} to estimate that only 1 in $\sim3030$~stars in our fields would exhibit variability detectable at the $5\sigma$ level over this timescale. Given that only $\sim4200$ stars were detected across all fields by DAOPHOT, we would only expect to detect $\sim 1$ variable star. Regarding extragalactic transients, based on supernova Ia rate for $R<21$~mag of 10 events per square degree per year \citep{Pain1996,Garnavich2004} and a 30\% fraction of SN Ia among local SNe \citep{Guillochon2016}, we estimate an 11\% probability of finding such an object across all our fields. Finally, our result is consistent with the LIGO detection of a binary black hole merger, for which no optical EM counterpart is expected.

\section{Summary}\label{conclusions}

The TOROS collaboration conducted a prompt search for the electromagnetic counterpart of the first gravitational-wave event reported by LIGO using the 1.5-m telescope of Estaci\'on Astrof\'{\i}sica Bosque Alegre (EABA) in C\'ordoba, Argentina. Our search spanned two nights, during which we targeted 21 fields containing 14 nearby ($D<60$~Mpc) galaxies with high probabilities of hosting the event. We covered $0.62\sq^{\circ}$ and reached a $5\sigma$ limiting AB magnitude of $r=21.7$. We used a combination of difference-imaging techniques and machine-learning procedures to detect and classify potential transients. No {\it bona fide} events were found, a result that is consistent with the low probability of detecting stellar or extragalactic variability given our temporal and areal coverage, and with the later classification of the GW event as a merger of two stellar-mass black holes.

Our host-galaxy ranking approach serves as a complementary strategy to the wide-field surveys for these transients, such as those conducted by the Dark Energy Survey \citep{Annis2016,Soares2016}, the intermediate Palomar Transient Factory \citep{Kasliwal2016},  MASTER \citep{2016arXiv160501607L}, Pan-STARSS \citep{2016arXiv160204156S}, and the VLT Survey Telescope (E. Brocato et al, in preparation). Given the incompleteness of local galaxy catalogs, the rapid dissemination of possible counterpart candidates by the wide-field surveys would enable detailed photometric coverage to be contributed by many modest-aperture, narrow-field telescopes throughout the world. Additionally, unfiltered CCD observations may be desirable at this stage given the large uncertainties in the possible colors of these counterparts.

\acknowledgments
{\it The TOROS collaboration acknowledges support from Ministerio de Ciencia, Tecnolog\'{\i}a e Innovaci\'on Productiva (MinCyT) and Consejo Nacional de Investigaciones Cient\'{\i}ficas y Tecnol\'ogicas (CONICET) from Argentina, grants from the National Science Foundation of the United States of America, NSF PHYS 1156600 and NSF HRD 1242090, and the government of Salta province in Argentina.}

\ \par

{\it Facilities:} \facility{EABA}.

\bibliographystyle{apj} 
\bibliography{toros}
\end{document}